\documentclass[preprint, amssymb,  aps, prl]{revtex4-1}

\usepackage{amsmath}
\usepackage{fullpage}
\usepackage{amssymb}
\usepackage{amsfonts}
\usepackage{color}
\usepackage[pdftex]{graphicx}
\usepackage{epstopdf}

\newcommand{\ephat}{\hat{\mathbf{e}}_p}
\newcommand{\zhat}{\hat{\mathbf{z}}}
\newcommand{\xhat}{\hat{\mathbf{x}}}
\newcommand{\yhat}{\hat{\mathbf{y}}}

\newcommand{\modk}{\vert \mathbf{k}_0 \vert }
\newcommand{\Mz}{M_{\left\{\zhat\right\}}}
\newcommand{\Rz}{R_z}

\newcommand{\Epq}{\mathbf{E}_{p,q}^{\mathbf{in}}}
\newcommand{\Eppq}{\mathbf{E}_{-p,-q}^{\mathbf{in}}}
\newcommand{\Tt}{\mathbf{T}}
\newcommand{\Etpq}{\mathbf{E}_{p,q}^{\mathbf{t}}}
\newcommand{\Etppq}{\mathbf{E}_{-p,-q}^{\mathbf{t}}}
\newcommand{\Sop}{\overline{\mathbf{S}}}
\newcommand{\epphat}{\hat{\mathbf{e}}_{-p}}
\newcommand{\Atpqp}{A_{p,q}^{\mathbf{t}}(x,y)}
\newcommand{\Btpqpp}{B_{p,q}^{\mathbf{t}}(x,y)}
\begin{document}

\title{Angular momentum-induced circular dichroism in non-chiral nanostructures}

\author{Xavier Zambrana-Puyalto$^{1,2}$, Xavier Vidal$^{1}$ and Gabriel Molina-Terriza$^{1,2} \email{gabriel.molina-terriza@mq.edu.au}$}

\affiliation{$^{1}$Department of Physics and Astronomy, Macquarie University, 2109 NSW, Australia \\ $^{2}$ARC Centre for Engineered Quantum Systems, Macquarie University, 2109 NSW, Australia}

\email{gabriel.molina-terriza@mq.edu.au}

\maketitle

{\bf{Circular dichroism (CD), i.e. the differential response of a system to left and right circularly polarized light, is one of the only techniques capable of providing morphological information of certain samples. In biology, for instance, CD spectroscopy is widely used to study the structure of proteins. More recently, it has also been used to characterize metamaterials and plasmonic structures. Typically, CD can only be observed in chiral objects. Here, we present experimental results showing that a non-chiral sample such as a sub-wavelength circular nano-aperture can produce giant CD when a vortex beam is used to excite it. These measurements can be understood by studying the symmetries of the sample and the total angular momentum that vortex beams carry. Our results show that CD can provide a wealth of information about the sample when combined with the control of the total angular momentum of the input field.}}

Since its discovery in the 19th century, circular dichroism (CD) has been widely used in science. Defined as the differential absorption of left and right circular polarization (LCP or RCP) \cite{Barron2004}, its uses are as diverse as protein spectroscopy, DNA studies and characterization of the electronic structure of samples \cite{Kelly2005}. In the advent of nano-photonic circuitry, a lot of work has been put recently into characterizing plasmonic components in terms of CD \cite{Decker2007,Hendry2012,Sersic2012}. Typically, it was thought that samples that produced CD had to be chiral, \textit{i.e.} they could not be superimposed with its mirror image \cite{Bishop2012}. However, recent experiments with planar plasmonic structures have shown that a non-zero CD can be obtained with a sample that lacks mirror symmetry with respect to the incident input beam \cite{Plum2009,Cao2012,Maoz2012,Ren2012}. These experiments show that mirror symmetric structures can show a non-null CD if the input beam is tilted with respect to the plane of the structure. Other attempts to create CD with a non-chiral sample have been carried out surrounding the sample with a chiral medium \cite{Abdulrahman2012,Xavi2013}. In this article, we show for the first time how to induce CD in a non-chiral sample, under normal incidence. In contrast to the previous approaches, the mirror symmetry of the system is broken with an internal degree of freedom of the input beam: its angular momentum (AM). The AM of light has gained a lot of interest since the seminal work of Allen and co-workers \cite{Allen1992}. One of the interesting properties of the AM of light is that photons carry AM in packets of $m \hbar$ units. Furthermore, it was shown that the AM of a beam is linearly related to the rotation speed that absorbing particles can achieve when they interact with the beam \cite{Friese1996}. Our experiments show that CD can be induced in a non-chiral sample if the two (left and right) circularly polarized modes are vortex beams. The reason behind this interesting phenomenon is that the input beams are not a mirror images of each other.


\begin{figure}[htbp]
\centering\includegraphics[width=13cm]{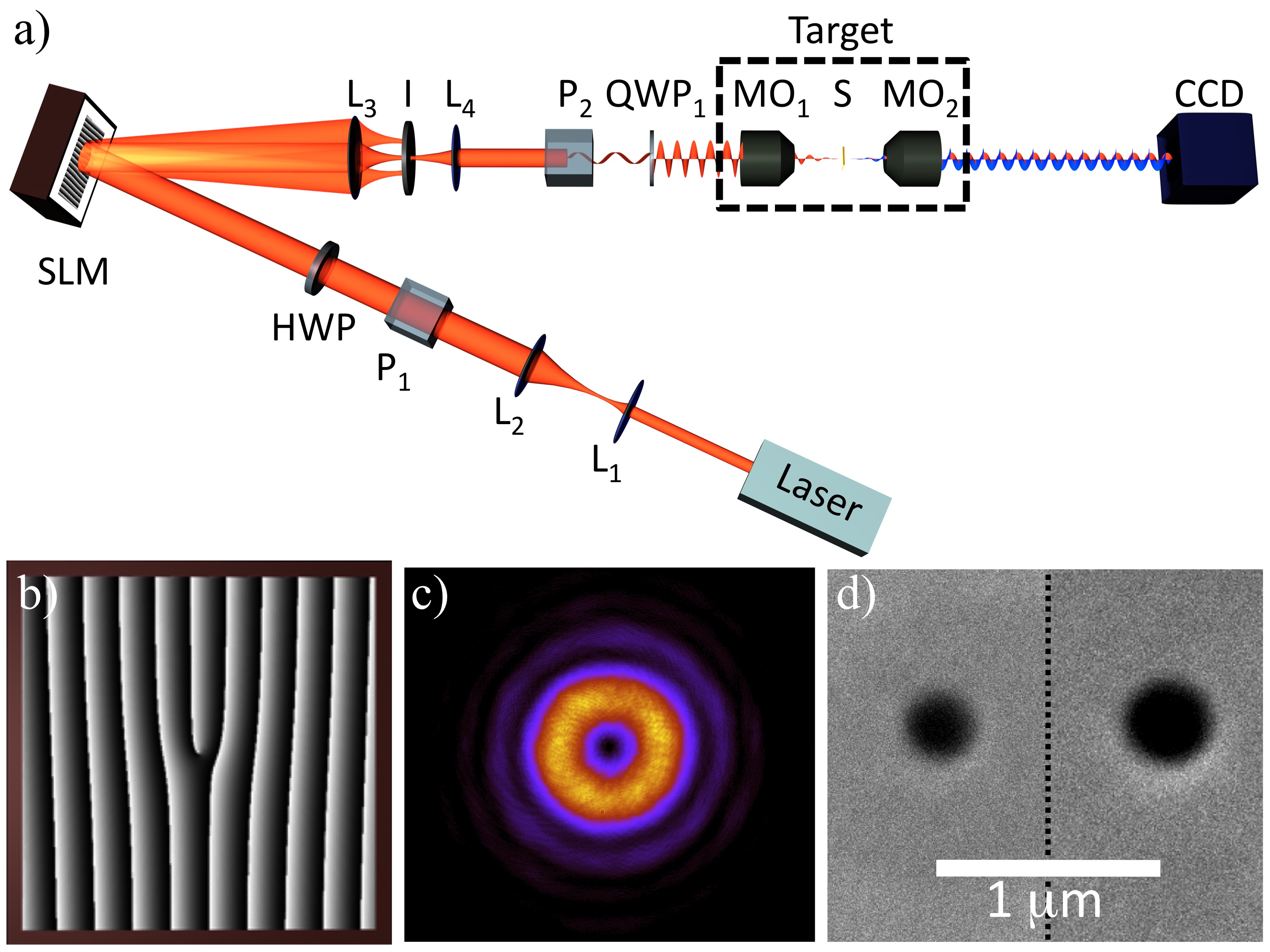}
\caption{(a) Schematic view of the optical set-up in consideration. A LCP and RCP vortex beam goes through a sub-wavelength circular aperture and their transmissivities are measured. First, we prepare an input Gaussian beam from a laser source in order to optimize the diffraction from the SLM. We expand it to match the size of the SLM chip with a telescope (lenses L$_1$-L$_2$) and then we control its polarization with a polarizer (P$_1$) and a half wave plate (HWP). A pitchfork-like hologram ((b)) is used to prepare a vortex beam ((c)) and the non-desired orders of diffraction are filtered with a lens (L$_3$) and an Iris (I). Lens L$_4$ is used to match the size of the back-aperture of the microscope objective. Then, we control its circular polarization with a second polarizer P$_2$ and a quarter wave plate QWP$_1$. After that, we focus the beam to the plasmonic structure (S) ((d)) with a microscope objective (MO$_1$) of 1.1 numerical aperture (NA). Finally, we collimate the scattered light from the sample with another microscope objective (MO$_2$) with NA$=0.9$ and measure the transmission with a charged-couple device (CCD) camera.}
\label{set-up}
\end{figure}

A sketch of the experimental set-up we used is depicted in Figure \ref{set-up}. It can be divided into three parts: preparation of states, non-paraxial interaction with the sample, and measurement. For the state preparation we use a CW laser working at wavelength $\lambda_0$=633 nm, producing a collimated, linearly polarized Gaussian beam. From this Gaussian beam, we create a vortex beam (see Figure \ref{set-up}(c))with a Spatial Light Modulator (SLM) by displaying an optimized pitchfork hologram \cite{Richard2011} (see Figure \ref{set-up}(b)). Proper control of the pitchfork hologram allows us to create a phase singularity of order $q$ in the center of the beam, i.e. the phase of the beam twists around its center from $0$ to $2 \pi q$ radians in one revolution. Note that when $q=0$, the SLM behaves simply as a mirror. We finish the preparation of the input beam by setting its polarization to either LCP or RCP. This change of polarization does not appreciably affect the spatial shape of the input beam. 

After this initial preparation, the light is focused down to interact with a plasmonic sample using a high numerical aperture (NA$=1.1$) microscope objective. The samples are circular nano-apertures drilled in a 200 nm gold film deposited on a 1 mm glass substrate (see Figure \ref{set-up}(d)). The diameters of the nano-apertures range from 200 to 450 nm (see Methods). They are centered with respect to the incident beam with a nano-positioning stage. The interaction of the light and the centered nano-aperture occurs in the non-paraxial regime. Typically, the nano-aperture only allows a small part of the incoming beam to be transmitted. The transmitted light is scattered in all directions. This facilitates the coupling of light to superficial modes of light, as it has been described theoretically and experimentally by many authors \cite{Solthesis,Ivan2011,Yi2012}. The transmitted light is then collected by another microscope objective. Finally, a camera is used to capture the transmitted intensity. 

The CD of our samples is measured using the following procedure: First, we create a vortex beam of optical charge $q$ with the SLM. Secondly, we rotate QWP$_1$ to polarize the beam with LCP. Then, we center the sample with respect to the beam, at the focal plane of MO$_1$. We measure the transmitted intensity $I_q^L$, where $L$ stands for the polarization of the beam (LCP) and $q$ for its optical charge. Then, we rotate QWP$_1$ again to change the polarization state to RCP and re-center the sample. Finally, we measure the transmitted intensity $I^R_q$. From there, the CD associated to the vortex beam with charge $q$ can be obtained in the usual manner: the transmitted intensity with LCP and RCP modes are subtracted and the result is normalized by their addition:
\begin{equation}
\text{CD}_q(\%)=\dfrac{I_q^L-I_q^R}{I_q^L+I_q^R} \cdot 100
\label{CD}
\end{equation} 

\begin{table}
\caption{\label{TabCD}Measurements of CD($\%$) for three different phase singularities $q$ as a function of the diameter of the nano-aperture. CD is computed using equation (\ref{CD}).}
\begin{ruledtabular}
\begin{tabular}{cccc}
 & $q=-1$& $q=0$ & $q=1$\\
\hline 
$d_1=237$ nm & $-78 \pm 6$ & $-0.130 \pm 0.003 $ & $68 \pm 6$  \\
$d_2=212$ nm &  $-91 \pm 6$ & $-1.843 \pm 0.004 $ & $84 \pm 6$ \\
\hline 
$d_3=325$ nm & $-10 \pm 4$ & $0.596 \pm 0.004 $ & $9.8 \pm 1.1$ \\
$d_4=317$ nm & $-7 \pm 5$ & $1.258 \pm 0.005 $ & $12 \pm 4$ \\
$d_5=333$ nm & $-9 \pm 4$ & $0.841 \pm 0.005 $ & $13 \pm 2$ \\
\hline
$d_6=432$ nm & $-22.3 \pm 0.8$ & $0.596 \pm 0.003 $ & $27.1 \pm 0.8$ \\
$d_7=429$ nm & $-34 \pm 2$ & $-1.055 \pm 0.007 $ & $37.5 \pm 1.3$ \\
$d_{8}=433$ nm & $-46.8 \pm 1.3$ & $0.629 \pm 0.004 $ & $45.9 \pm 0.8$ \\
\end{tabular}
\end{ruledtabular}
\end{table}
\begin{figure}[htbp]
\centering\includegraphics[width=11cm]{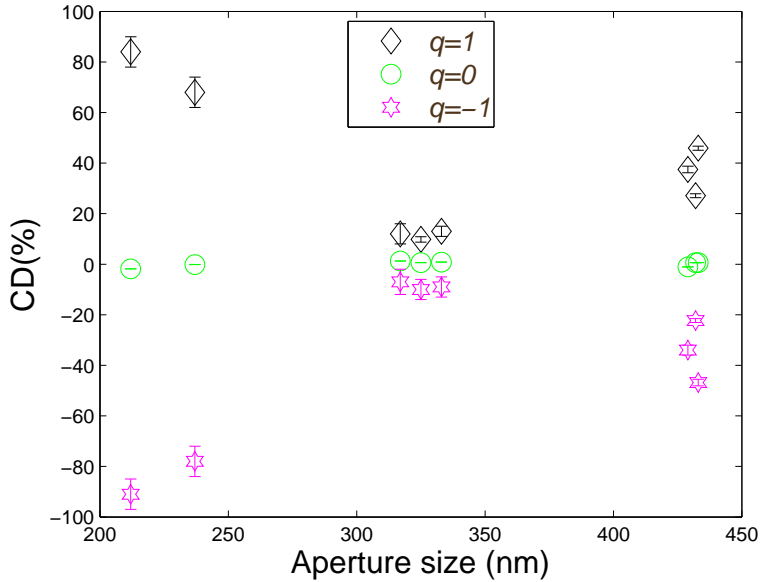}
\caption{Measurements of CD($\%$) for three different phase singularities $q$ as a function of the diameter of the nano-aperture. CD is computed using equation (\ref{CD}). Black diamonds are used to represent the CD obtained with $q=1$; green circles for $q=0$ and magenta hexagon stars for $q=-1$.}
\label{CD_fig}
\end{figure}
Our results are presented in Table \ref{TabCD} and Figure \ref{CD_fig}. The first column of Table \ref{TabCD} shows the size of the nano-aperture in consideration. The rest of columns show the measured CD using beams with different phase singularities of order $q=-1,0,1,$ respectively. As expected, due to the fact that the circular nano-aperture is mirror symmetric and the incidence is normal, the CD is very close to zero when the incident beam is Gaussian ($q=0$). The residual CD can be attributed to small asymmetries on the sample or the incoming beam. When vortex beams with $q=-1,1$ are used the situation is very different. We obtain a very large value for the CD, of the order of $90\%$ for some nano-apertures, even though the incidence is normal and the nano-aperture is still mirror symmetric. Furthermore, it can be observed that there is an underlying symmetry relating the value of CD$_1$ and CD$_{-1}$. Indeed, CD$_1 \simeq -$CD$_{-1}$. 

Let us now discuss these results. In order to understand the appearance of CD in a circular sample under normal incidence, we have to take a careful look at the symmetries of the system and the light probing the samples. We can start considering the symmetries of the target $\Tt$ comprising the two microscope objectives MO$_1$, MO$_2$ and the sample S (see Figure \ref{set-up}(a)). This will simplify the discussion, as we will just consider the interaction of a paraxial beam with the $\Tt$, as well as the output beams which will also be paraxial. Then, as both the circular nano-aperture and the microscope objectives have cylindrical symmetry along an axis normal to them, $\Tt$ is cylindrically symmetric. Without loss of generality we will label the symmetry axis as $z$ and we will denote the rotations around this axis by $\Rz$. Furthermore, $\Tt$ is also symmetric under any mirror transformation that contains the $z$ axis, \textit{e.g.} a transformation flipping the $x$ axis and leaving the other axis invariant. We will refer to such transformations as $\Mz$.

Now we will turn our attention to the symmetries of the light beams. The electric field of the light beam incident on $\Tt$ can be described within the paraxial approximation with the complex vector:
\begin{equation}
\Epq= A\rho^q e^{\left( iq\phi + i \modk z \right)} e^{(-\rho^2/w_0)} \ephat
\label{Epq}
\end{equation}
where $\ephat = (\xhat + i p \yhat) / \sqrt{2}$, $\xhat$ and $\yhat$ being the horizontal and vertical polarization vectors, $A$ is a normalisation constant, $\modk=2 \pi / \lambda_0$ with $\lambda_0$ the wavelength in consideration, and $\rho$, $\phi$, $z$ are the cylindrical coordinates. An implicit harmonic $\exp(-i 2\pi c t/\lambda_0)$ dependence is assumed, where $c$ is the speed of light. Note that $p=1$ refers to LCP and $p=-1$ to RCP. Now, if we apply a rotation around the $z$ axis by an angle $\theta$ to this beam, the resulting beam acquires a constant phase: $\Rz(\theta) \Epq =\exp(-i (p+q) \theta) \Epq$. This kind of beams are then eigenstates of the generator of rotations around the $z$ axis, \textit{i.e.} the $z$ component of the total angular momentum $J_z$: $J_z \Epq = (p+q) \Epq$. The $J_z$ eigenvalue is $m=p+q$. Also note that a beam with $p=0$ would be linearly polarized and will be no longer an eigenstate of $J_z$, since a rotation of the field would not leave it invariant. In fact, a closer look to equation (\ref{Epq}) enables us to see that when $p=\pm 1$ the beams $\Epq$ are eigenstates of the helicity operator $\Lambda$ with value $p$ \cite{Tung1985,Ivan2012PRA,Ivan2013}. That is, $\Lambda \Epq = p \Epq$, where $\Lambda = \textbf{J}\cdot\textbf{P}/|\textbf{P}|$ and $\mathbf{P}$ is the linear momentum operator. Hence, with our notation, a beam $\Epq$ with $p=1$ is both LCP and also an eigenvector of $\Lambda$ with value $1$, whereas a beam $\Epq$ with $p=-1$ is RCP and its helicity equals to $-1$. We would like to emphasize that this simple relation between helicity and polarization is only valid because, as mentioned previously, we are describing the system in the paraxial regime. When the paraxial approximation does not hold, the polarization of the field and helicity can no longer be so simply related \cite{Ivan2012PRA}. Now, if we apply a mirror transformation to the incident beam we obtain:
\begin{equation}
\Mz \Epq = \exp(i \alpha) \Eppq. 
\label{-Epq}
\end{equation}
where $\alpha$ is a phase given by the specific mirror transformation chosen. That is, we obtain a beam with the angular momentum and the helicity eigenvalues flipped. This is a consequence of the fact that both $J_z$ and $\Lambda$ anticommute with the mirror symmetry transformations $\Mz$ \cite{Messiah1999}: $M^{\dagger}_{\left\{\zhat\right\}} J_z \Mz = - J_z$ and $M^{\dagger}_{\left\{\zhat\right\}} \Lambda \Mz = - \Lambda$.

We will now explain our experimental results. We will start by classifying the transmitted electric field $\Etpq$ with the parameters $p$ and $q$ from the incident field: $p=-1,1$ and $q=-1,0,1$. Remember that for the incident field $\Epq$, $p$ is modified with QWP$_1$ and $q$ with the SLM (see Figure \ref{set-up}). Mathematically, the field $\Etpq$ can be obtained through the use of a linear operator, $\Sop$, which can be found using the Green dyadic formalism and contains all the relevant information about target $\Tt$ \cite{Martin1998,Ivan2011}. That is, $\Etpq=\Sop \{\Epq\}$, where the action of $\Sop$ on the incident field will in general be in the form of a convolution. As $\Sop$ is simply the mathematical description of $\Tt$, $\Sop$ inherits the symmetries of $\Tt$. Thus, due to the cylindrical and mirror symmetries of the target and given an incident field $\Epq$, the following statements hold: First, the transmitted field $\Etpq$ will also be an eigenstate of $J_z$ with the same eigenvalue of $m=p+q$ (see Methods). Second, two incident beams which are mirror images of each other will produce two transmitted fields which will be mirror images (see Methods). That is, given two mirror symmetric beams such as $\Epq$ and $\Eppq$ (see equation (\ref{-Epq})), their transmitted beams $\Etpq=\Sop \{\Epq\}$ and $\Etppq=\Sop \{\Eppq\}$ will be connected via a mirror symmetry: $\Etpq=\exp(i\alpha)\Mz\mathbf{E}_{-p,-q}^{\mathbf{t}}$. The proof can be found in Methods, but the physical idea is the following one: The invariance of the system under mirror transformations links the output of mirror inverted inputs. This last result is the key point to understand our CD measurements with vortex beams presented in Table (\ref{TabCD}) and Fig. (\ref{CD_fig}). In equation (\ref{CD}), the intensities can be obtained from the transmitted electric field:
\begin{equation}
I_q^{L/R}=\int_{-\infty}^{\infty} \int_{-\infty}^{\infty} |\mathbf{E}_{+1/-1,q}^{\mathbf{t}}|^2 {dx} {dy},
\label{inten}
\end{equation}
where the integral is taken on the plane of the detector (in our case CCD chip of the camera). Then, for a mirror symmetric sample, it can be proven (see Methods) that
\begin{equation}
I_q^{L/R}=I_{-q}^{R/L}
\label{Iq}
\end{equation}
Let us apply it to prove that $CD_0=0$ and that $CD_q=-CD_{-q}$. When $q=0$, we obtain that $I_0^L=I_{0}^R$. Substituting this in the definition of CD, equation (\ref{CD}), gives us $CD_0=0$. However, when $q\neq0$, equation (\ref{Iq}) leads us to $I_q^L-I_q^R = I_{-q}^R - I_{-q}^L = - (I_{-q}^L-I_{-q}^R)$, which implies that $CD_q=CD_{-q}$, in very good agreement with our measurements. 

\begin{figure}[htbp]
\centering\includegraphics[width=14cm]{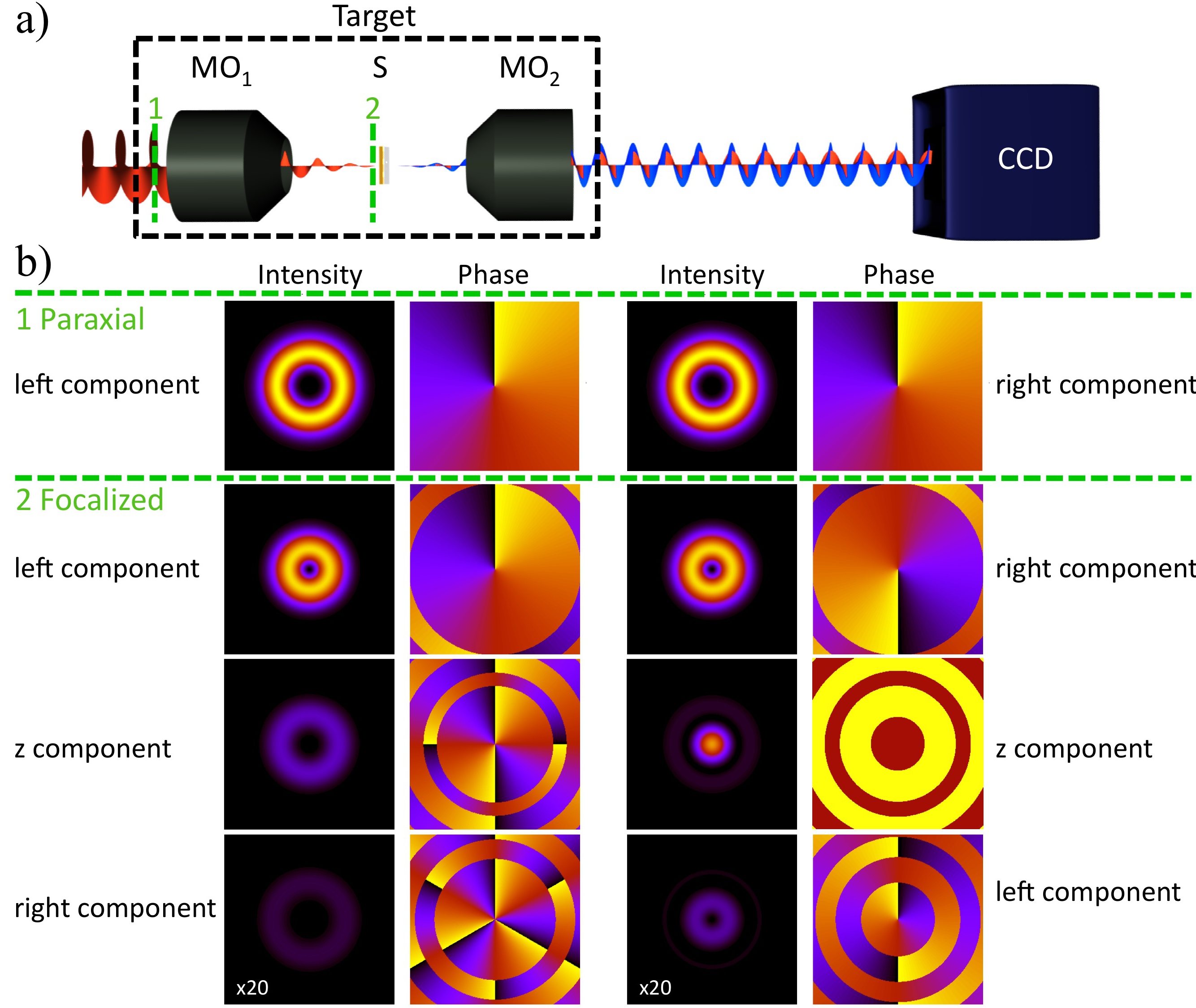}
\caption{\textbf{a)} Measurement part of the set-up. A paraxial beam $\mathbf{E}_{p,1}^{\mathbf{in}}$ is focused by MO$_1$ onto the sample. Then, MO$_2$ collimates the transmitted light and $\mathbf{E}_{p,1}^{\mathbf{t}}$ is imaged recorded by a CCD camera. \textbf{b)}Intensity and phase plots of the two beams used to carry out the measurement of CD$_{q=1}$, \textit{i.e.} $\mathbf{E}_{1,1}^{\mathbf{in}}$ and $\mathbf{E}_{-1,1}^{\mathbf{in}}$. In the upper row, the intensity and phase of the beams are shown on the back-aperture plane of the MO$_1$. Both the intensity and the phase can be described with eq \ref{Epq}. In contrast, the three rows below show the intensity and the phase of the same beams ($\mathbf{E}_{1,1}^{\mathbf{in}}$ and $\mathbf{E}_{-1,1}^{\mathbf{in}}$) at the focal plane of MO$_1$. As it can be observed, even though their paraxial intensities and phases are analogous, their structure is completely different at the focal plane. This is a direct consequence of the fact that the AM of both beams differ in two units.}
\label{Projection}
\end{figure}

Powerful as these symmetry considerations are, they still cannot explain the quantitative results of CD we obtain, nor their variation with the diameter of the nano-aperture. Note that in our experiments, there are large variations of the vortex-induced CD for holes around 220 nm, where the CD reaches around 90\%, as compared with the holes with sizes around 320 nm, which present a CD around 10\%. The symmetry arguments only indicate that when $q\neq0$ the two opposite circular polarizations are not the mirror image of each other and then the associated CD$_q$ does not have to be zero. Here, the AM of light plays a crucial role again. In general, it can be observed that CD measurements compare the differential ratio of electromagnetic fields with opposite circular polarization and a difference of AM of 2 units. For example, CD$_{q=1}$ relates $\vert \mathbf{E}_{1,1}^{\mathbf{t}} \vert^2$ and $\vert \mathbf{E}_{-1,1}^{\mathbf{t}} \vert^2 $, whose respective AM values are $m_{p=1}=1+1=2$ and $m_{p=-1}=1-1=0$. It is then interesting to observe that CD can also be sensitive to differential absorption of AM states. In the case of the nano-aperture, this is the most probable cause of the giant value of CD obtained in the experiments. Even though the sample is cylindrically symmetric, thus preserving the AM of field, input beams with different values of AM have very different scattering amplitudes. This is very similar  to what happens in the scattering from spherical objects, where different spherical modes (the so called multipolar modes) are scattered with different amplitudes (the Mie coefficients). This problem can be analytically studied using the Generalized Lorenz-Mie Theory \cite{Bohren1983,GLMT_book}. Using the model of the aplanatic lens \cite{WolfII1959,Novotny2006}, one can check that the fields at the focal plane of MO$_1$ produced by $\mathbf{E}_{-1,q}^{\mathbf{in}}$ and $\mathbf{E}_{1,q}^{\mathbf{in}}$ are very different (see Figure \ref{Projection}(b)) - and consequently, so are their multipolar decompositions \cite{Gabi2008,Zambrana2012}. These two different focused fields couple very differently to the multipolar moments of the structure. In the case of a spherical object, the coupling of multipolar modes depends drastically on its diameter. Furthermore, the relation between the coupling coefficient and the sphere radius is non-linear and rather complex \cite{Zambrana2012}, giving rise to the so-called Mie resonances in the scattering of the sphere. Then, given two spheres of different sizes, different coupling scenarios can occur not only depending on the geometry of the particles but also on the beams used to excite these particles. Very similar effects have been experimentally observed for other plasmonic structures \cite{Yuri2008,Banzer2010SRR,Rodriguez-Herrera2010}. Then, CD with vortex beams can potentially provide useful information about the differential absorption or scattering of AM modes. 


In conclusion, we have observed a giant CD on a subwavelength circular aperture induced by vortex beams. We studied the transmision of beams with three different phase singularities of order $q=-1,0,1$. We have seen that the results can be conveyed from a symmetry perspective. In particular, we have proved that, even for mirror symmetric systems, CD can be induced if the LCP and RCP beams are not connected via a mirror symmetry. We show that the information carried by CD measurements has two different contributions: the differential scattering of different circular polarization states but also the differential scattering of different angular momentum states. Thus, we expect that vortex beam-induced circular dichroism will be able to unveil properties of the sample which are hidden to the standard circular dichroism measurements. Finally, it is interesting to see that in other related phenomena, such as molecular optical activity, the interplay between the two symmetries associated to helicity and AM (electromagnetic duality and rotational symmetry) are also essential to fully understand the problem from first principles \cite{Ivan2013JCP}. 

{\bf Methods} \\
\small
We consider the rotation transformations ($R_z$), mirror transformations ($\Mz$) and angular momentum ($J_z$) operators, as well as the scattering operator ($\Sop$) as linear integro-differential operators acting on the electric vector of the complex electromagnetic field.

\textbf{Preservation of $J_z$ for the transmitted field}\\
Due to the invariance of $\Tt$ under rotations around the $z$ axis, the linear operator $\Sop$, which contains all the information about the system, commutes with the rotations $\Rz$: $R^{\dagger}_z \Sop \Rz =  \Sop $. Due to the bijective properties of the exponential function, the same commutation relation holds for the generator of rotations, $J_z$: $J^{\dagger}_z \Sop J_z =  \Sop $. Then, given that $\Epq$ is an eigenvector of $J_z$, $\Etpq$ must also be an eigenvector of $J_z$ with the same eigenvalue:
\begin{equation}
J_z \Etpq=J_z \Sop\{\Epq\}=\Sop J_z \{\Epq \} = (p+q) \Sop \{\Epq\} = (p+q) \Etpq
\end{equation}
\textbf{Scattering of mirror symmetric beams} \\
Due to the invariance of $\Tt$ under mirror symmetries, the linear operator $\Sop$, which contains all the information about the system, commutes with $\Mz$: $M^{\dagger}_{\{ \zhat\}} \Sop \Mz =  \Sop $. Now, given two mirror symmetric beams such as $\Epq$ and $\Eppq$ (see equation (\ref{-Epq})), it can be checked that their respective transmitted fields ($\Etpq$ and $\Etppq$) are related with a mirror symmetry:
\begin{equation}
\Etpq=\Sop \{ \Epq\} = \Sop \{ \exp(i\alpha) \Mz \Eppq \} = \Mz \Sop \{ \exp (i \alpha)\Eppq \} = \exp(i \alpha) \Mz \Etppq
\label{mirror}
\end{equation}
\\
\textbf{Proof of} $\bf{I_q^{L/R}=I_{-q}^{R/L}}$\\
As mentioned in the body of the manuscript, the target is cylindrically symmetric scatterer. Hence, if the incident field $\Epq$ is an eigenstate of $J_z$ with eigenvalue $(p+q)$, then the transmitted field $\Etpq$ needs to remain an eigenstate of $J_z$ with the same eigenvalue $(p+q)$. Nevertheless, the helicity is, in general, not preserved in the interaction. This phenomenon is a consequence of the duality symmetry being highly broken by the nano-aperture and the multilayer system \cite{Ivan2012PRA,Ivan2013,Zambrana2013OE,Bliokh2013}. Because duality symmetry is highly broken by the sample, the helicity of the incident beam $\Epq$ is not preserved. Thus, the field $\Etpq$ comprises two helicity components, one of polarization $\ephat$ and another one of polarization $\epphat$: 
\begin{equation}
\Etpq = \Atpqp\ephat  + \Btpqpp\epphat
\label{decomp}
\end{equation}
where $\Atpqp$ and $\Btpqpp$ are the complex amplitudes of the two polarizations at the plane of the camera. We call $\Atpqp$ the direct component maintains the polarization state $\ephat$. The other orthogonal component is $\Btpqpp$, and we call it crossed component. The crossed component has a polarization state $\epphat$ when the incident state is $\ephat$. Now, we can use the definition of $I_q^{L/R}$ on equation (\ref{inten}), the decomposition of $\Etpq$ in its two orthogonal components given by equation (\ref{decomp}), and the fact that $\ephat^* \cdot \epphat = 0$ to obtain that $I_q^{L/R}$ can be expressed as:
\begin{equation}
I_q^{L/R} = \int_{-\infty}^{\infty} \int_{-\infty}^{\infty} |A_{+1/-1,q}^{\mathbf{t}}(x,y)|^2 + |B_{+1/-1,q}^{\mathbf{t}}(x,y)|^2 {dx} {dy}
\label{iq}
\end{equation}
Following an identical procedure, the following equation yields for $I_{-q}^{R/L}$:
\begin{equation}
I_{-q}^{R/L} = \int_{-\infty}^{\infty} \int_{-\infty}^{\infty} |A_{-1/+1,-q}^{\mathbf{t}}(x,y)|^2 + |B_{-1/+1,-q}^{\mathbf{t}}(x,y)|^2 {dx} {dy}
\label{i-q}
\end{equation}
Choosing the mirror symmetric operator to be $\Mz = M_{\{x \rightarrow -x \}}$, then we can use equation (\ref{-Epq}) to get a relation between the coefficients in equations (\ref{iq}, \ref{i-q}):
\begin{equation}
\begin{array}{ccl}
&& \Etpq  = \Atpqp\ephat  + \Btpqpp\epphat  =  \Mz \Etppq = \\
&& = \Mz \left( A_{-p,-q}^{\mathbf{t}}(x,y) \epphat + B_{-p,-q}^{\mathbf{t}}(x,y) \ephat  \right)  =  - A_{-p,-q}^{\mathbf{t}}(-x,y) \ephat + B_{-p,-q}^{\mathbf{t}}(-x,y) \epphat 
\end{array}
\end{equation} 
which implies that $\Atpqp = - A_{-p,-q}^{\mathbf{t}}(-x,y)$ and $\Btpqpp = B_{-p,-q}^{\mathbf{t}}(-x,y)  $, due to the orthogonality of $\ephat$ and $\epphat$. Consequently, it follows that:
\begin{eqnarray}
I_q^{L/R} = \int_{-\infty}^{\infty} \int_{-\infty}^{\infty} |A_{+1/-1,q}^{\mathbf{t}}(x,y)|^2 + |B_{+1/-1,q}^{\mathbf{t}}(x,y)|^2 {dx} {dy} = \nonumber \\
= \int_{-\infty}^{\infty} \int_{-\infty}^{\infty} |- A_{-1/+1,q}^{\mathbf{t}}(-x,y)|^2 + |B_{-1/+1,q}^{\mathbf{t}}(-x,y)|^2 {dx} {dy} =  \\ 
= \int_{-\infty}^{\infty} \int_{-\infty}^{\infty} |A_{-1/+1,-q}^{\mathbf{t}}(x^\prime,y)|^2 + |B_{-1/+1,-q}^{\mathbf{t}}(x^\prime,y)|^2 {dx^\prime} {dy} = I_{-q}^{R/L} \nonumber
\end{eqnarray}
as the integrations limits remain the same under the change $x \rightarrow -x^\prime$. \\
\textbf{Fabrication of samples}\\
The tested nanoholes were fabricated by milling with a FIB on a gold layer of 200 nm, deposited on top of a 1 mm thick glass substrate. The distance between them is 50 $\mu$m, thus avoiding the coupling of surface plasmons launched from one nanohole to the closest neighbor.\\
\textbf{Imaging of the samples}\\
The images were taken with a secondary electron Scanning Electron Microscope (SEM, JEOL JSM-6480) operated at 10KeV. The images were analyzed with Matlab where the boundaries of the nanoholes were determined by selecting the pixels whose intensity was below the 10$\%$ of the maximum. The presented images on Figure \ref{set-up} were not post-processed. The obtained diameters are listed in Table \ref{TabCD} on the manuscript.\\
\textbf{Nanopositioning system}\\
The sample is mounted on a piezo electric transducer (PZT) on closed-loop with resolution below 0.5nm (translation range 300 $\mu$m with 20-bit USB interface and noise floor of tens of picometers). \\

{\normalsize \bf Acknowledgments\\}
\small The authors want to thank Mathieu L. Juan for helpful advice while setting up the experiment and Alexander E. Minovich for the preparation of the sample. The sample was prepared at the ACT Node of the Australian Nanofabrication Facility, and the SEM images were taken at the Microscopy Unit of Faculty of Science in Macquarie University. This work was funded by the Australian Research Council Discovery Project DP110103697 and the Centre of Excellence in Engineered Quantum Systems (EQuS). G.M.-T. is the recipient of an Australian Research Council Future Fellowship (project number FT110100924).

{\normalsize \bf Author Contributions\\}
\small XZP and XV set up the experiment and did the measurements. XZP and GMT wrote the manuscript. GMT supervised the project. All the authors participated in the analysis of the results.

{\normalsize \bf Competing financial interests\\}
\small The authors declare no competing financial interests

\end{document}